\documentclass[twocolumn]{aastex63}

\def\Mc{\ensuremath{\mathcal{M}_{\rm c}}}
\def\chieff{\ensuremath{\chi_\mathrm{eff}}}

\def\q{\ensuremath{q}}

\def\BF{\ensuremath{\mathcal{B}}}
\def\OR{\ensuremath{\mathcal{O}}}

\def\CHE{\textsc{CHE}}

\usepackage{color}

\shortauthors{Qin et al.}
\usepackage{lineno}
\begin{document}

\title{Searching for candidates of coalescing binary black holes formed through chemically homogeneous evolution in GWTC-3}

\correspondingauthor{Yuan-Zhu\, Wang; Simone S. Bavera; Ying\, Qin}
\email{wangyz@pmo.ac.cn; Simone.Bavera@unige.ch; yingqin2013@hotmail.com}

\author[0000-0002-2956-8367]{Ying\, Qin}
\affiliation{Department of Physics, Anhui Normal University, Wuhu, Anhui 241000, China}

\author[0000-0001-9626-9319]{Yuan-Zhu\, Wang}
\affiliation{Key Laboratory of Dark Matter and Space Astronomy, Purple Mountain Observatory, Chinese Academy of Sciences, Nanjing, 210033, China}

\author[0000-0002-3439-0321]{Simone\,S.\,Bavera}
\affiliation{Département d’Astronomie, Université de Genève, Chemin Pegasi 51, CH-1290 Versoix, Switzerland}
\affiliation{Gravitational Wave Science Center (GWSC), Université de Genève, CH-1211 Geneva, Switzerland}

\author[0000-0002-9188-5435]{Shichao\, Wu}
\affiliation{Max-Planck-Institut f{\"u}r Gravitationsphysik (Albert-Einstein-Institut), D-30167 Hannover, Germany}
\affiliation{Leibniz Universit{\"a}t Hannover, D-30167 Hannover, Germany}

\author{Georges\, Meynet}
\affiliation{Département d’Astronomie, Université de Genève, Chemin Pegasi 51, CH-1290 Versoix, Switzerland}
\affiliation{Gravitational Wave Science Center (GWSC), Université de Genève, CH-1211 Geneva, Switzerland}

\author{Yi-Ying\, Wang}
\affiliation{Key Laboratory of Dark Matter and Space Astronomy, Purple Mountain Observatory, Chinese Academy of Sciences, Nanjing, 210033, People's Republic of China}
\affiliation{School of Astronomy and Space Science, University of Science and Technology of China, Hefei, Anhui 230026, People's Republic of China}

\author[0000-0002-6442-7850]{Rui-Chong Hu}
\affiliation{Guangxi Key Laboratory for Relativistic Astrophysics, School of Physical Science and Technology, Guangxi University, Nanning 530004, China}

\author[0000-0002-9195-4904]{Jin-Ping Zhu}
\affiliation{Department of Astronomy, School of Physics, Peking University, Beijing 100871, China} 

\author[0000-0001-9424-3721]{Dong-Hong\, Wu}
\affiliation{Department of Physics, Anhui Normal University, Wuhu, Anhui 241000, China}

\author{Xin-Wen\, Shu}
\affiliation{Department of Physics, Anhui Normal University, Wuhu, Anhui 241000, China}

\author{Fang-Kun\, Peng}
\affiliation{Department of Physics, Anhui Normal University, Wuhu, Anhui 241000, China}

\author{Han-Feng\, Song}
\affiliation{College of Physics, Guizhou University, Guiyang city, Guizhou Province, 550025, P.R. China}

\author{Da-Ming\, Wei}
\affiliation{Key Laboratory of Dark Matter and Space Astronomy, Purple Mountain Observatory, Chinese Academy of Sciences, Nanjing, 210033, People's Republic of China}
\affiliation{School of Astronomy and Space Science, University of Science and Technology of China, Hefei, Anhui 230026, People's Republic of China}

\begin{abstract}
The LIGO, Virgo, and KAGRA (LVK) collaboration has announced 90 coalescing binary black holes (BBHs) with $p_{\rm astro} > 50\%$ to date, however, the origin of their formation channels is still an open scientific question. Given various properties of BBHs (BH component masses and individual spins) inferred using the default priors by the LVK, independent groups have been trying to explain the formation of the BBHs with different formation channels. Of all formation scenarios, the chemically homogeneous evolution (CHE) channel has stood out with distinguishing features, namely, nearly-equal component masses and preferentially high individual spins aligned with the orbital angular momentum. We perform Bayesian inference on the BBH events officially reported in GWTC-3 with astrophysically-predicted priors representing different formation channels of the isolated binary evolution (CEE: common-envelope evolution channel; CHE; SMT: stable mass transfer). Given assumed models, we report strong evidence for GW190517\_055101 being most likely to have formed through the CHE channel. Assuming the BBH events in the subsample are all formed through one of the isolated binary evolution channels, we obtain the lower limits on the local merger rate density of these channels at $11.45 ~\mathrm{Gpc^{-3}~yr^{-1}}$ (CEE), $0.18 ~\mathrm{Gpc^{-3}~yr^{-1}}$ (CHE), and $0.63 ~\mathrm{Gpc^{-3}~yr^{-1}}$ (SMT) at $90\%$ credible level.
\end{abstract}


\section{Introduction}
To date, a total of 90 merging binary black holes (BBHs) with $p_{\rm astro}$ \footnote{$p_{\rm astro}$, the probability that a gravitational wave event is of astrophysical origin, is defined by comparing the different rates of gravitational waves and background events at a given value of a search statistic \citep{2015PhRvD..91b3005F,2016PhRvX...6d1015A,2018ApJ...861L..24L}.} of at least 50\%, have been reported in the third LIGO, Virgo, and KAGRA (LVK) Collaboration Gravitational-Wave Transient Catalog \citep[GWTC-3,][]{2021arXiv211103634T,2021arXiv211103606T}. GW190517\_055101, first reported in GWTC-2 \citep{2021PhRvX..11b1053A} with $p_{\rm astro} > 0.99$, has a primary BH mass $M_1 = 37.4^{+11.7}_{-7.6} M_{\odot}$ and a secondary BH mass $M_2 = 25.5^{+7.0}_{-7.3} M_{\odot}$ (mass ratio $q = M_2 / M_1 \sim 0.68$) at the 90\% credible level. This event has stood out with the most highest effective spin in GWTC-3 \footnote{Note that GW190403\_051519 has a lower $p_{\rm astro}$ ($p_{\rm astro} > 50\%)$, although it was reported with the $\chi_{\rm eff}$ = $0.70^{+0.15}_{-0.27}$.} ($\chi_{\rm eff} = 0.52^{+0.19}_{-0.19}$). At the leading post-Newtonian order, the gravitational wave signals are largely dependent on the effective spin parameter $\chi_{\rm eff}$ \citep{2001PhRvD..64l4013D}, which is defined as:
\begin{equation}\label{xeff} 
    \chi_{\rm eff} = \frac{(M_1\vec{\chi}_1+M_2\vec{\chi}_2)\cdot \hat{L}}{M_1+M_2} ,
\end{equation}
where $M_1$ and $M_2$ are the component masses, $\vec{\chi}_1$ and $\vec{\chi}_2$ are their corresponding dimensionless spin parameters, and $\hat{L}$ is the unit vector along the orbital angular momentum (AM).

Various formation channels of BBHs \citep[see recent reviews of][]{2022PhR...955....1M,2021hgwa.bookE..16M} have been proposed since the first discovery of gravitational waves, GW150914 \citep{2016PhRvL.116f1102A}. The leading formation channels can be divided into two broad categories, i.e., isolated binary evolution and dynamical formation. The former category includes, (i) classic isolated binary evolution scenario involving a common-envelope evolution phase \citep[CEE channel, e.g.,][]{1993MNRAS.260..675T,1997MNRAS.288..245L,1998ApJ...506..780B,2002ApJ...572..407B,2007PhR...442...75K,2016Natur.534..512B,2020A&A...635A..97B,2022ApJ...933...86Z}, (ii) a double stable mass-transfer phase \citep[SMT channel, e.g.,][]{2017MNRAS.471.4256V,2017MNRAS.468.5020I,2019MNRAS.490.3740N,2021A&A...647A.153B,2021ApJ...921L...2O,2022A&A...660A..26B} or both stars evolving chemically homogeneously \citep[CHE channel, e.g.,][]{2016MNRAS.458.2634M,2016A&A...588A..50M,2016MNRAS.460.3545D,2020MNRAS.499.5941D,2021MNRAS.505..663R,2021ApJ...910..152Z,2022A&A...657L...8B}. Dynamical formation in dense stellar environments includes formation in globular clusters, young stellar clusters and open stellar clusters \citep[e.g.,][]{2015PhRvL.115e1101R,2018PhRvL.121p1103F}, active galactic nuclei disks \citep{2018ApJ...866...66M,2020ApJ...899...26T}, and isolated triple or higher-order stellar systems \citep[e.g.,][]{2017ApJ...836...39S,2018ApJ...863....7R,2020PhRvD.101j4053G,2020A&A...640A..16T}. 

In general, the effective spin $\chi_{\rm eff}$ has been widely regarded as a discriminator for disentangling the isolated (individual BH spins preferentially aligned to the direction of the orbital AM) and dynamical (random orientations of individual BH spins) formation scenarios \citep{2016ApJ...832L...2R,2017Natur.548..426F,2018ApJ...854L...9F,2017PhRvD..96b3012T,2017CQGra..34cLT01V,2017MNRAS.471.2801S}. Of all the formation channels mentioned above, the CHE channel has been considered as the most likely scenario that can lead to the formation of BBH systems with: (i) nearly equal masses \citep{2016MNRAS.458.2634M,2016A&A...588A..50M}, (ii) two BHs with preferentially high spins that are aligned to the orbital AM \citep{2016A&A...588A..50M,2020MNRAS.499.5941D,2021ApJ...910..152Z,2022A&A...657L...8B}. Therefore, these two features can be used as a probe to investigate whether or not there is quantitative high-confidence evidence of any merging BBHs formed through the CHE channel. Recent studies \citep{2021PhRvD.104h3010R,2021ApJ...921L..15G,2022arXiv220906978V} have pointed out that the inferred spin and mass parameters are dependent on the choice of priors \footnote{Prior: probability distribution that represents knowledge or uncertainty of a data object before observing it.}. This is because the measurements of spin and mass are poorly constrained, and so the resulting broad posteriors can be heavily swayed by one's choice of prior. In particular, some specific events have also been recently reanalysed with different priors than the official LVK analysis. Assuming as a prior that the more massive BH has a zero-spin and that the rotation axis of the less massive BH is aligned with the orbital AM, \cite{2020ApJ...895L..28M} argued that, in the context of isolated binary evolution, the less massive component of GW190412 could be highly spinning. \cite{2020ApJ...899L..17Z} further investigated how the choice of a prior can influence parameter estimates of GW190412. \citet{2021ApJ...922L..14M} suggested that a prior of nonspinning BH for GW200115 is more consistent with current astrophysical understanding. \cite{2020ApJ...904L..26F} pointed out that GW190521 likely straddles the pair-instability gap by reanalyzing its signal with a population-informed prior on less massive BH mass. These findings confirm that the choice of a prior can play a critical role in inferring the properties of gravitational-wave sources.

Different groups have independently investigated the formation channels of observed BBHs \citep[e.g.,][]{2021PhRvD.103h3021W,2022arXiv220604062W,2021MNRAS.507.5224B,2022MNRAS.511.5797M,2022PhRvD.105h3526F}. Recently, \citet{2021ApJ...910..152Z} found that multiple channels are required when interpreting the currently released LVK's BBHs, assuming a limited number of model uncertainties and a subsample of all possible formation scenarios. In their models which are publicly available, the predictions for spin and mass distributions have been presented for various astrophysical formation channels of BBHs. In this work, we perform Bayesian inference to search for evidence of BBHs most likely formed through the CHE channel by considering the models of isolated binary evolution CEE/SMT/CHE  of \citet{2021A&A...647A.153B} and \citet{2020MNRAS.499.5941D} \citep[as released by][]{michael_zevin_2021_4947741} as the astrophysically-predicted priors. In Section \ref{sec:2}, we first briefly introduce the CHE and its predicted properties of BBHs. Then we present our Bayesian analysis and results in Section \ref{sec:3}. Finally, the main conclusions and some discussion are summarized in Section \ref{sec:4}.

\section{Properties of BBHs predicted by the CHE channel} \label{sec:2}
Chemical mixing induced by fast rotation leads to massive stars evolving homogeneously, without expanding to become a red supergiant star \citep{1987A&A...178..159M}. \cite{2013A&A...554A..23M} performed a spectroscopic analysis of Wolf-Rayet stars in the Large Magellanic Cloud and Milky Way Galaxy, and found that some of these objects might have gone through the CHE. In order to sustain efficient mixing throughout their lifetimes, single massive stars must be rotating quickly at birth, requiring metal-poor environments, where the stellar winds are weak \citep{2001A&A...369..574V}. For massive stars that can be efficiently spun up by the tidal interaction in close binaries at subsolar metallicities, rotationally enhanced mixing has been predicted to produce the CHE for both stars \citep{2016MNRAS.458.2634M,2016A&A...588A..50M,2016A&A...585A.120S} or only the more masssive component \citep{2009A&A...497..243D,2017A&A...604A..55M,2019ApJ...870L..18Q}. Observations for the O-type stars in six nearby Galactic open stellar clusters show that $\sim$ 70\% of O-type stars are in close binaries and about 1/3 of them are able to interact on the main sequence \citep{2012Sci...337..444S}. Furthermore, \cite{2013ApJ...764..166D} simulated a massive binary-star population to find that the rapid rotation of massive stars could be obtained via mass transfer or mergers. Alternatively, the CHE induced by the accretion from its companion could also be reached for massive stars with weak tidal interactions in relatively wide orbits \citep{2007A&A...465L..29C}. Recently, \cite{2022arXiv220803999G} claimed that the accretion-induced CHE could be an important formation channel producing electromagnetic transients like GRBs/Ic-BL (SLSN-I/Ic-BL)\footnote{GRB: gamma-ray burst; Ic-BL: Broad-lined Type Ic Supernova; SLSN: superluminous Supernova type I} under the collapsar (magnetar) scenario. In addition to CEE and SMT, the CHE was also found to play a critical role in contributing to the long GRB rate \citep{2022A&A...657L...8B}.

Massive stars with nearly equal masses in close binaries tend to follow the CHE and thus result in binary BHs that could merge within a Hubble time \citep{2016MNRAS.458.2634M,2016A&A...588A..50M,2016MNRAS.460.3545D,2020MNRAS.499.5941D,2021MNRAS.505..663R,2021ApJ...910..152Z,2022arXiv220402619B,2022A&A...657L...8B}. Treatments of CHE and predictions of its outcome have been calculated by a variety of techniques, ranging from simplified prescriptions \citep[e.g.,][]{2016MNRAS.458.2634M,2021MNRAS.505..663R} to detailed models of massive binary evolution \cite{2016A&A...588A..50M,2020MNRAS.499.5941D}. As mentioned earlier, current models predict that BBHs formed through the CHE are expected to have mass-ratios close to unity and preferentially high inspiral effective spins $\chi_{\rm eff}$. First, \cite{2016MNRAS.458.2634M} showed that there is a strong preference for two BHs with comparable masses, and especially that there are no BH binaries of interest with the mass ratio $q < 0.5$. Therefore, the mass ratio $q = 0.5$ can be used as a lower limit for forming merging BBHs through the CHE channel. Additionally, \cite{2016A&A...588A..50M,2020MNRAS.499.5941D} found that BBHs originated from the CHE tend to have mass ratio $q > 0.8$. Second, as expected for BBHs formed through {\CHE}, high BH spins can be reached \citep[see Fig. 9 in][]{2016A&A...588A..50M}. We would expect even higher effective spins $\chi_{\rm eff}$ for BBHs formed through the CHE if less efficient AM transport within their progenitors is assumed \citep[e.g., see the green line for $\chi_{\rm eff}$ in the bottom panels of Fig. 1 in][]{2021ApJ...910..152Z}.

\cite{2021ApJ...921L...2O} used population synthesis models to find that, two equal-mass helium stars might be formed after the CE phase and then produce two fast-spinning BHs with the mass ratio $q = 1$. Such double helium-star systems, however, are more common to be produced through the CHE channel \citep[see Fig. 3 for a parameter-space study of detailed binary calculations in][]{2016A&A...588A..50M}, which most likely produce two equal-mass BHs. Therefore, in this work, we consider the BBHs with equal masses to be more likely formed through the CHE channel. For BBHs formed via the SMT channel, the orbital separation after the second mass transfer phase is much wider when compared with CEE. Therefore, the $\chi_{\rm eff}$ was expected to be very low, i.e., $\chi_{\rm eff}$ $< 0.1$ \citep[see Fig. 2 in][]{2021A&A...647A.153B}, assuming the accretion onto BHs is Eddington-limited\footnote{BHs can be efficiently spun up in binaries if one assumes that hypercritical accretion \cite{2022RAA....22c5023Q} or mildly super-Eddington accretion \cite{2022ApJ...930...26S} is allowed.}. For the channel of the CEE \citep[see detailed investigations of this channel in][]{2020A&A...635A..97B}, it has been recently shown in \cite{2022ApJ...924..129Q} that the upper limit of $\chi_{\rm eff}$ is constrained to be not higher than 0.5, assuming that the first-born BH is formed from an initially more massive star with a strongly efficient AM transport mechanism (i.e., the revised version of original Tayler-Spruit dynamo in \cite{2019MNRAS.485.3661F}, similar results were also yielded in \cite{2018A&A...616A..28Q,2020A&A...636A.104B} and \cite{2022MNRAS.511.3951F} for a recent investigation\footnote{The spins of BHs born from single stars have been predicted to be small ($\chi$ $\lesssim$ 0.01, \cite{2018A&A...616A..28Q,2019ApJ...881L...1F}, and $\chi$ $\sim 0.1$, \cite{2020A&A...636A.104B}).}). 

It is important to note that identifying events from different channels requires a solid understanding of the predictions of these channels. To date, the predictions for the different channels are plagued due to major uncertainties \citep[e.g., ][]{2022ApJ...925...69B,2022MNRAS.516.5737B}. In this work, we are focused on searching for possible candidates that could have very similar properties predicted by the CHE channel under our current understanding of its relevant physical processes.

\begin{deluxetable*}{@{\extracolsep{4pt}}c ccccc ccc}
\label{tab:BF}
\tablecaption{Logarithmic odd ratios ($\ln \OR$) and Bayes factors ($\ln \BF$) of the CHE prior and the SMT prior compared to the CEE prior.}
\tablehead{
\colhead{Name} & \colhead{CHE}& \colhead{SMT}\\
& $\ln \OR (\ln \BF)$ & $\ln \OR (\ln \BF)$
}
\startdata
GW190517\_055101 & $3.4(4.2)$ & $-8.7(-9.3)$ \\
GW190805\_211137 & $2.4(3.1)$ & $-0.7(-1.4)$ \\
GW190620\_030421 & $1.4(2.1)$ & $-2.7(-3.3)$ \\
GW170729 & $1.2(1.9)$ & $-1.4(-2.1)$ \\
GW200216\_220804 & $0.5(1.2)$ & $0.9(0.2)$ \\
GW190719\_215514 & $-0.6(0.1)$ & $0.3(-0.4)$ \\
GW190527\_092055 & $-2.1(-1.4)$ & $1.3(0.6)$ \\
GW200128\_022011 & $-2.1(-1.4)$ & $2.3(1.6)$ \\
GW170823 & $-3.2(-2.5)$ & $1.8(1.1)$ \\
GW190513\_205428 & $-3.4(-2.7)$ & $0.7(0.0)$ \\
GW190727\_060333 & $-3.5(-2.7)$ & $1.9(1.2)$ \\
GW190828\_063405 & $-4.3(-3.6)$ & $0.7(0.1)$ \\
GW170809 & $-6.0(-5.2)$ & $1.7(1.0)$ \\
GW190630\_185205 & $-7.1(-6.3)$ & $1.0(0.4)$ \\
GW200224\_222234 & $-7.4(-6.7)$ & $2.4(1.8)$ \\
GW200129\_065458 & $-12.2(-11.5)$ & $1.3(0.6)$ \\
GW170814 & $-14.9(-14.2)$ & $2.0(1.3)$ \\
GW190925\_232845 & $-16.3(-15.6)$ & $-4.0(-4.7)$ \\
GW191103\_012549 & $-18.6(-17.9)$ & $-16.1(-16.8)$ \\
GW190930\_133541 & $-20.0(-19.3)$ & $-18.5(-19.2)$ \\
GW200316\_215756 & $-21.8(-21.1)$ & $-20.9(-21.6)$ \\
GW151226 & $-41.6(-40.9)$ & $-40.3(-41.0)$ \\
GW190728\_064510 & $-46.9(-46.1)$ & $-48.3(-48.9)$ \\
GW191129\_134029 & $-57.6(-56.9)$ & $-18.8(-19.5)$ \\
GW191204\_171526 & $-106.8(-106.0)$ & $-21.8(-22.5)$ \\
GW191216\_213338 & $-124.4(-123.7)$ & $-127.4(-128.1)$ \\
\enddata
\end{deluxetable*}

\section{Analysis and Results} \label{sec:3}
\begin{figure*}[htbp!]
     \centering
     \includegraphics[width=\textwidth]{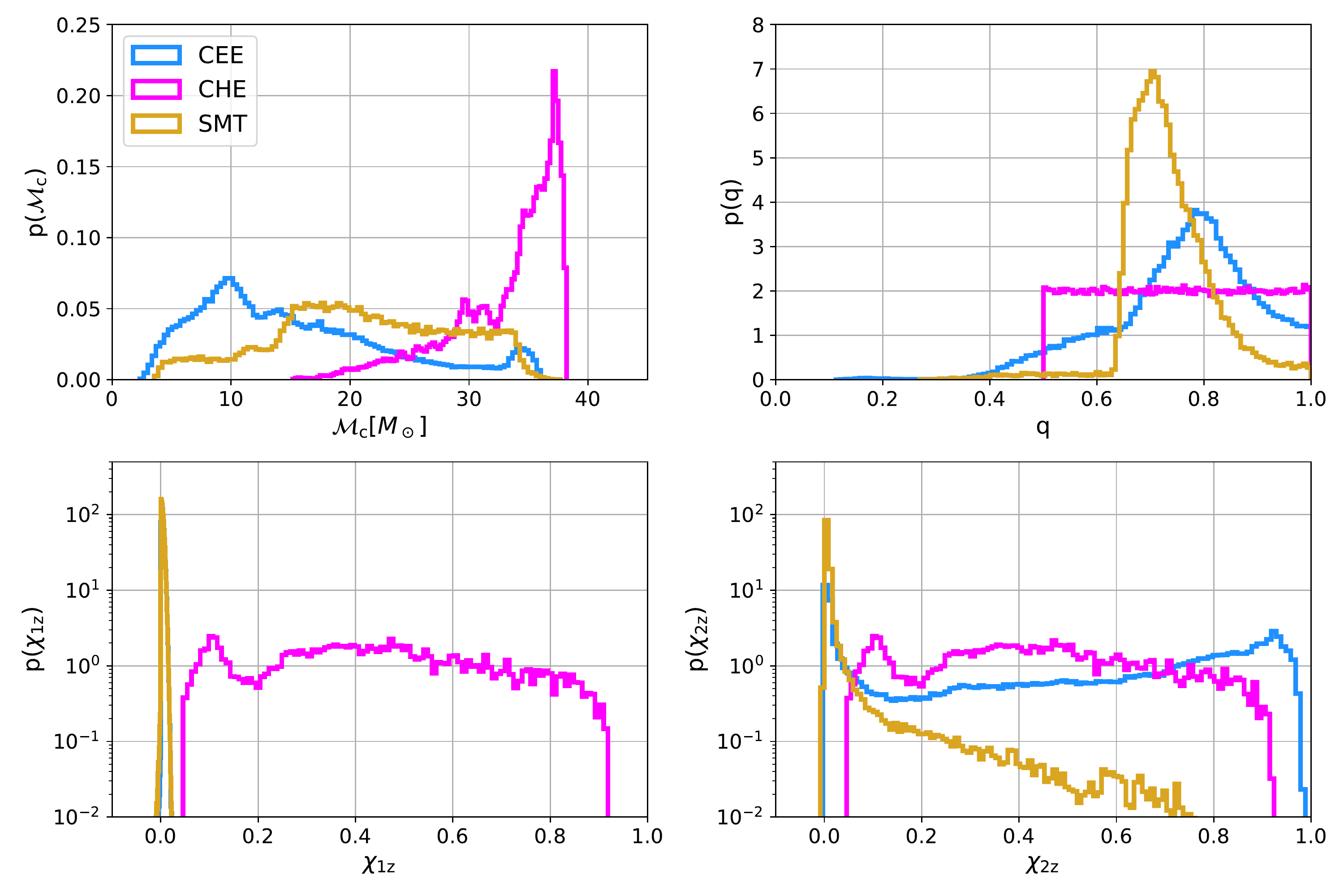}
     \caption{Theoretical predictions of the marginal distributions (blue: CEE; magenta: CHE; green: SMT) of chirp mass (${\Mc}$), mass ratio ($q$) and component spins ($\chi_{1z} = \chi_{1}, \chi_{2z} = \chi_{2}$, assuming both $\chi_{1z}$ and $\chi_{1z}$ are aligned to the direction of the orbital AM}). The simulated data are obtained from the publicly released models in \citet{2021ApJ...910..152Z}, except that for the CHE channel we extend the $q$ distribution down to 0.5, which is in agreement with the studies in \cite{2016MNRAS.458.2634M,2016MNRAS.460.3545D}.
     \label{fig:priors}
\end{figure*}

We perform Bayesian inference on the BBH events with various priors representing different formation channels of the isolated binary evolution (CEE, CHE, and SMT). Fig.~\ref{fig:priors} shows the model predictions \citep[we refer the interested readers to the detailed descriptions in ][]{2021ApJ...910..152Z} on the marginal distributions of chirp mass ${\Mc}$\footnote{${\Mc} = \frac{(M_1 M_2)^{3/5}}{(M_1 + M_2)^{1/5}}$, where $M_1$ and $M_2$ are the masses of the two BHs, respectively.}, mass ratio $q$ and component spins ($\chi_1, \chi_2$) of different BBH formation channels of the isolated binary evolution. Ideally, one should use the joint distributions ($p(\Mc,q,\chi_1, \chi_2)$) as priors in the inference to preserve the potential correlations as predicted by the models. However, it is numerically difficult to approximate the 4-D probability distribution function (PDF) with a kernel density estimation (KDE), so for simplicity we adopt 1D marginal distributions as priors in this study. For the CHE channel, we extend the lower limit of the mass ratio $q$ down to 0.5 (0.8 adopted in \cite{2021ApJ...910..152Z}, cf. Fig.~\ref{fig:priors}) to be consistent with the predictions of \cite{2016MNRAS.458.2634M}, assuming the same spin values  ($\chi_1 = \chi_2$) for the two BHs formed through the CHE channel.  

We carry out a preliminary selection for the potential candidates being formed through CHE. The events listed in Tab.~\ref{tab:BF} are candidates that pass our selection criteria. They have properties (inferred by the LVK's default prior) that satisfied: i) $\mathcal{M}_{\rm c,M} < 40~M_\odot$; ii) $q_{\rm M} > 0.5$; iii) $\chi_{\rm 1z,M} > 0.05$, where the subscribe ``${\rm M}$" represents the median value of the posterior samples.

The Bayesian inference takes the strain data of an event, the waveform model, the power spectral density, and the prior for the parameters representing the binary's properties as input, and returns the parameters' posterior distributions as output. One can derive the Bayes factor, \BF, between two models (waveform $+$ prior) by comparing the Bayesian evidence $\mathcal{E}$ \citep{2019PASA...36...10T,2020ApJ...899L..17Z}. The Bayes factor reflects a comprehensive evaluation of the goodness-of-fit and the prior volume for the two models. Therefore, larger is the Bayes factor, more favored is the formation channel indicated.

We choose the source frame chirp mass {\Mc}, the mass ratio $q$, the individual BH spin magnitude $\chi_i$ and the tilt angle of component spin $\theta_i$ to describe the intrinsic properties of the BBHs in the sampling. Although several individual BBHs have been identified as possibly exhibiting precessional effects due to spin misalignment \citep{2020PhRvD.102d3015A,2021PhRvD.103j4027I,2022PhRvD.106b3019H,2022PhRvD.106b4009C,2022arXiv220313406V,2022ApJ...924...79E,2022arXiv220611932P}, such evidence for individual-event precessional has generally been weak or inconclusive. In this study, since BH spins are expected to be aligned with the orbital AM vector in the CHE channel, we adopt $\theta_i=0$ in our inference by assuming aligned spins in all of our samples. To increase the efficiency of the parameter estimation process, we also make a simplification of the priors of $\chi_1$: we fix $\chi_1=0$ for the CEE and SMT channels, since for these two channels we assume all first-born BHs have negligible $\chi_{\rm 1z}$ ($\chi_{\rm 1z}$ $<0.01$) as shown in Fig.~\ref{fig:priors}. 

We use the publicly available data for the selected events from the Gravitational Wave Open Science Center\footnote{\url{https://www.gw-openscience.org/eventapi/html}}, and adopt a duration of 17 seconds covering the detection time in the analysis. The results are produced using the python package \textit{bilby} \citep{2019ascl.soft01011A} and its built-in sampler \textit{dynesty} \citep{2020MNRAS.493.3132S}. The low and high frequency cutoffs for the likelihood calculation are set to 20 (30) Hz and 512 Hz respectively for LIGO (Virgo) detectors. We employ the Phenom waveform family in the inference, and choose the model IMRPhenomXP \citep{2021PhRvD.103j4056P} since it achieves a good trade-off between accuracy and speed.

\begin{figure*}[htbp!]
     \centering
     \includegraphics[width=\textwidth]{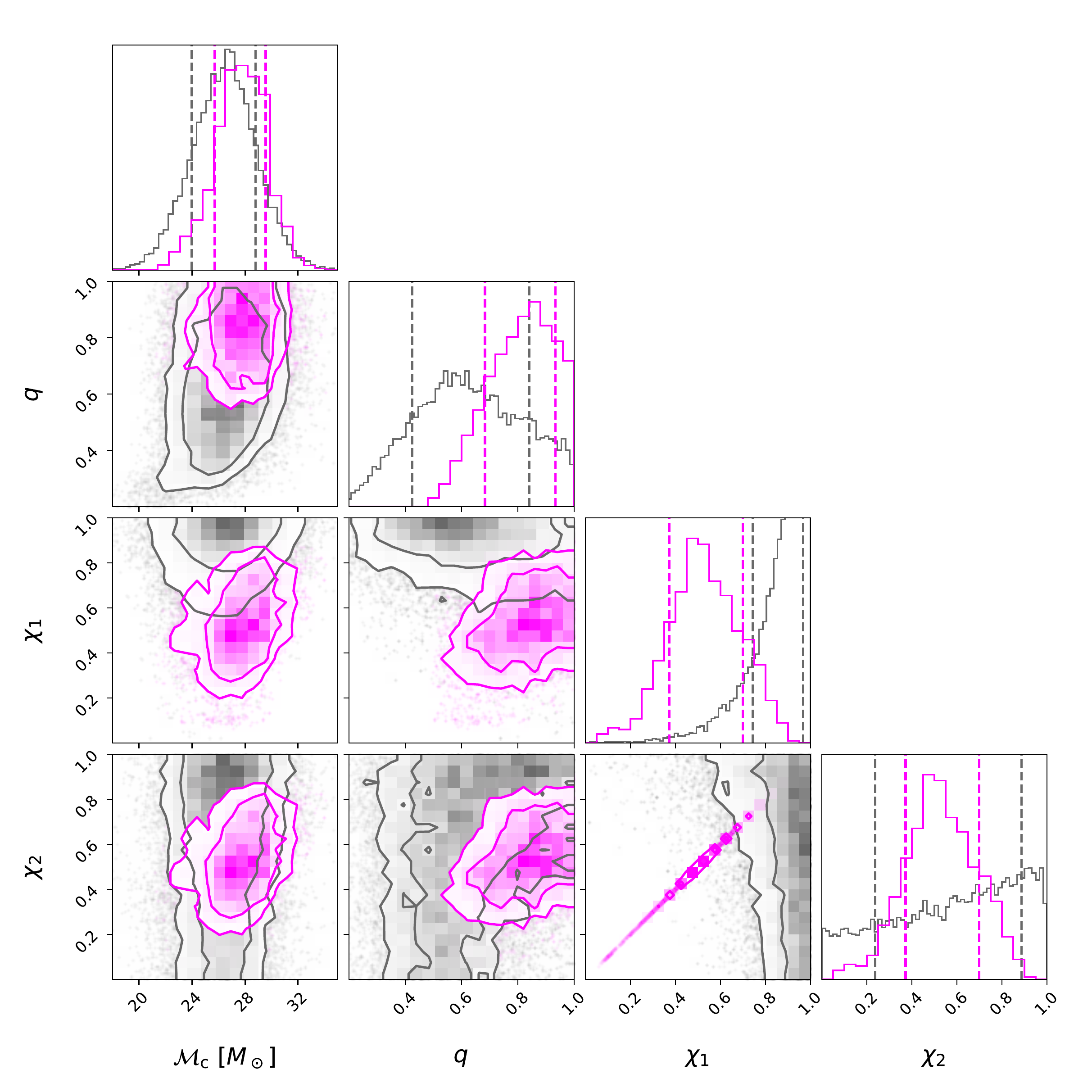}
     \caption{The posterior distributions for GW190517\_055101 inferred by LVK's default prior (gray) and the CHE prior (magenta). The inner and outer solid circles in the 2-D plots mark the $68\%$ and $90\%$ credible regions, respectively.}
     \label{fig:190517}
\end{figure*} 

The resulting Bayes factors of the CHE prior as well as the SMT prior compared to the CEE prior are listed in the brackets in Tab.~\ref{tab:BF}. In addition to \BF, we also consider the theoretical predictions of the merger rate density in the local Universe (z = 0) of each channel as prior odds to compute the odds ratio (\OR): $\OR = \BF \times (\mathcal{R}_i/\mathcal{R}_{\rm CEE})$, where $\mathcal{R}_i/\mathcal{R}_{\rm CEE}$ is the relative rate of a specific channel with respect to the CEE channel. We adopt the predictions in \cite{2022A&A...657L...8B}, showing the CEE, CHE and SMT channel would contribute to $29\%$, $14\%$ and $57\%$ of the events originated from isolated binary evolution. The adopted mixing fractions are consistent with the models used in \cite{2021ApJ...910..152Z}. In practice these values might change due to model uncertainties \citep{2022LRR....25....1M}. We use the odds ratio as the final indicator to determine the preference for formation channels. If {\OR} is larger than 3/30/100 \citep{2021ApJ...913L...7A}, one can conclude that a moderate/strong/very strong evidence of preference is found for a prior over the CEE prior. On the other hand, if $1/3 < \OR <3$, the priors have comparable support from the data. In Tab.~\ref{tab:BF}, we find that 6 out of 26 events have $\OR > 1/3$ ($\ln \OR > -1.1$) for the CHE prior, while 14 events have $\OR > 1/3$ for the SMT prior. We also note that several events show a strong favor for the CEE prior (see the bottom in Tab.~\ref{tab:BF}). This result could be explained by adopting combined priors of chirp mass, mass ratio, and two spin components, especially for $\chi_{\rm 1z}$ $\sim$ 0 and a more broad distribution of $\chi_{\rm 2z}$ (i.e., from 0 to 1).

Most notably, for the CHE prior, GW190517\_055101 has the largest {\OR} of $\ln \OR = 3.4$ ($\OR = 30$), which is at the threshold of being strong evidence for the preference of CHE prior against the CEE prior. These results may be due to the fact that the CHE prior allows for much larger individual spins when compared with the CEE and SMT channels, which is needed to better match the gravitational waveform for this event; in addition, the 1-D chirp mass prior for the CHE channel has larger probability densities around the inferred chirp mass $\sim 28 M_\odot$. In contrast, the logarithmic {\OR} for the SMT prior is $-8.7$, indicating that this prior is much less supported than the others. Additionally, the finding also indicates that the CEE prior is not favored. This is most likely due to the lower $\chi_{\rm 1z}$ predicted by the two channels (CHE and SMT). Assuming BBHs originated from isolated binary evolution and given the considered models, our analysis favors the CHE origin of GW190517\_055101. In Fig.\ref{fig:190517}, we show the properties of this event inferred by LVK's default prior \citep{2021PhRvX..11b1053A}, and the result inferred with our CHE prior (magenta) for comparison. Under the default prior, both BH individual spins have most of their posterior supports at $\chi_i \sim 1$ (though $\chi_2$ is barely constrained). The marginal mass ratio distribution peaks at $q \sim 0.6$, and there is still posterior supports at $q = 1$. Despite a nearly identical constraint on the chirp mass, using our CHE prior yields different posteriors for $q$, $\chi_1$, and $\chi_2$. The individual spin distributions peak at $\chi_i \sim 0.5$; extreme values ($\chi_i \sim 0$ and $\chi_i \sim 1$) for the spins are also excluded. The peak of the mass ratio distribution shifts to $q \sim 0.9$, and the distribution has a low probability density at the lower edge (0.5) of the mass ratio prior. \citet{2021ApJ...922L...5C} recently showed that the mass ratio of GW190517\_055101 under a population-informed prior can be relatively low (down to $\sim 0.25$) assuming it follows the $\chieff-q$ anti-correlation, which was also corroborated by \cite{2022arXiv220803405A} using a different statistical approach. This confirms that different prior assumptions may lead to diverse results and the CHE prior in this work provides an alternative solution to the parameter space of $q$ for this event. Other events, GW170729, GW190620\_030421, and GW190805\_211137 have moderate support on the CHE prior.

\begin{figure*}[htbp!]
     \centering
     \includegraphics[width=\textwidth]{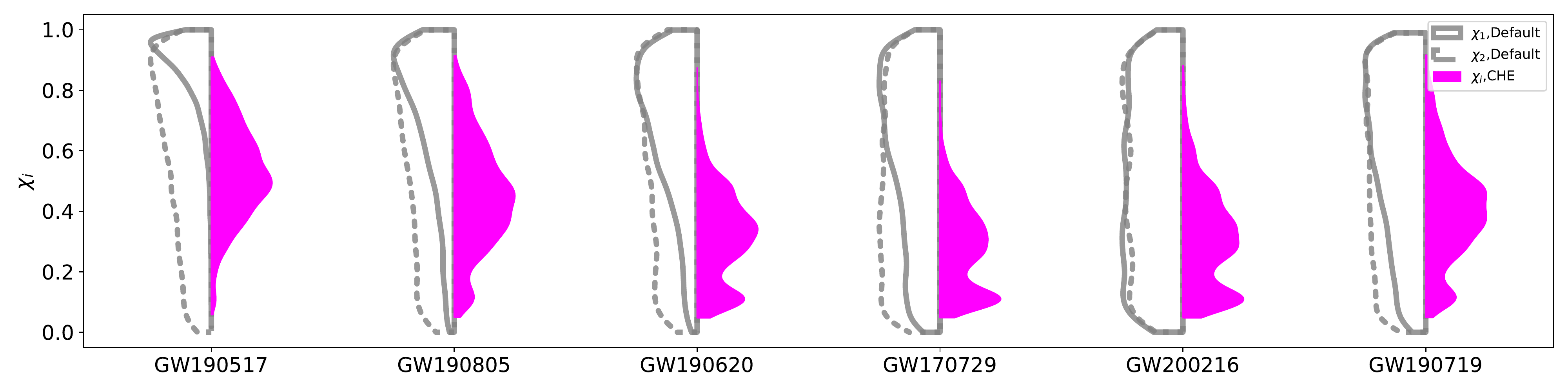}
     \includegraphics[width=\textwidth]{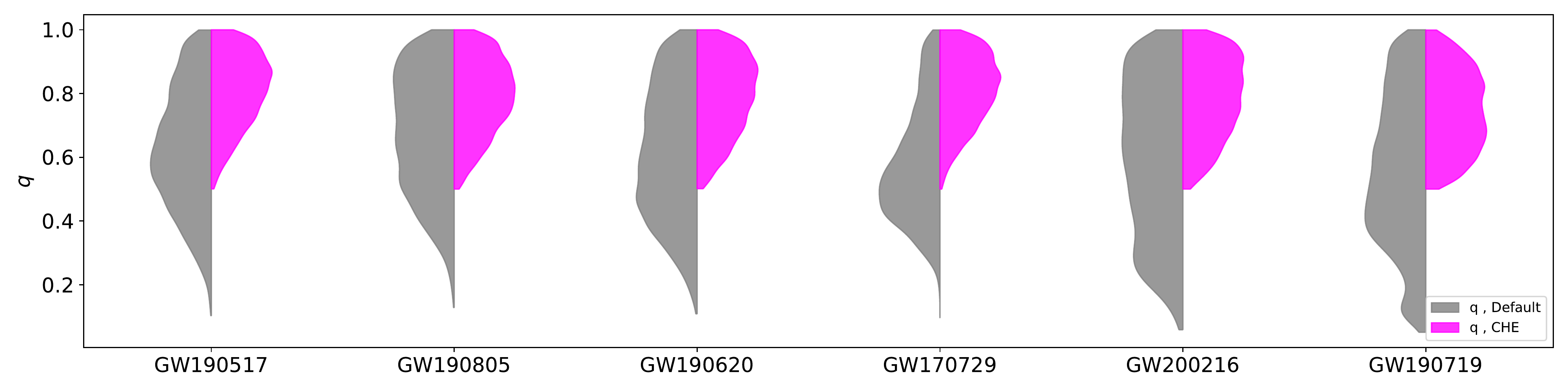}
     \caption{Violin plot showing marginal posterior distributions of individual spins (upper panel) and mass ratios (lower panel) inferred with LVK's default prior and the CHE prior. These selected events have odd ratios larger than $1/3$ for the CHE channel. For brevity, we omit the suffix of the event name.}
     \label{fig:violin}
\end{figure*}

In Fig.~\ref{fig:violin}, we show posterior distributions of the six events with $\OR > 1/3$ on the CHE channel to illustrate the impact of the prior. Similar to GW190517\_055101, the mass ratio distributions of the other events all peak at $q \gtrsim 0.8$, and their posterior supports drop rapidly towards the lower edge of the prior. On the other hand, by comparing the distributions in Fig.~\ref{fig:violin} with the priors as shown in Fig.~\ref{fig:priors}, one can observe that our choices of prior strongly affect the posterior distribution of $\chi_i$ and $q$ compared to the LVK 's default prior. We also show the results for the eight events that most favor ($\OR > 3$) the SMT hypothesis in Appendix.~\ref{sec:sfig}.

With the Bayes factors and posterior distributions of each event, one can derive the astrophysical merger rate density ($\mathcal{R}0$) in the local Universe ($z=0$) for each channel. Following the method described in \citet{kim2003,2016PhRvX...6d1015A}, we calculate the ``event-based" merger rate density for each event. The method treats every single event as an unique subclass of BBHs and calculates their merger rate density separately. Then, the total event rate is the sum of the individual rates. The Poisson fluctuation and the uncertainties of the parameters inferred with each prior are taken into account in the calculation (see Sec.~VI and Appendix C in \citet{2016PhRvX...6d1015A} for details). The injection provided by \citet{2021arXiv211103634T} is utilized to estimate the instruments' sensitivity for searching mergers similar to each event. We consider an evolving merger rate density with the form $\mathcal{R}_i = \mathcal{R}0_i(1+z)^{2.7}$, where $\mathcal{R}0_i$ is the local merger rate density for the BBHs with properties similar to the $i$-th event in our sample. Adopting a uniform prior for the local merger rate density, we obtain the posterior distributions for $\mathcal{R}0_i$ of different events. Second, we randomly draw a value from the $\mathcal{R}0_i$ posterior distribution for each event, and multiply each of them by a weight decided by the events' {\BF}s (for a particular event, the sum of weights for the three channels is 1). Then, we sum over these re-weighted values to obtain the overall $\mathcal{R}0$s. By repeating the above steps 50,000 times, we numerically derived the probability distributions for $\mathcal{R}0$ of each channel. Since our selected events make up a subset of the entire catalog, the derived results can only be regarded as the lower limits for $\mathcal{R}0$. By integrating the probability distributions, the lower limits are $11.45 ~\mathrm{Gpc^{-3}~yr^{-1}}$, $0.18 ~\mathrm{Gpc^{-3}~yr^{-1}}$ and $0.63 ~\mathrm{Gpc^{-3}~yr^{-1}}$ at 90\% credible level for the CEE, CHE and SMT channel, respectively.

\section{Conclusions and Discussion} \label{sec:4}
Although significant progress in gravitational-wave astrophysics today has been made in the past several years, the origin of the BBHs remains an open scientific question. In the modeling for the origin of BBHs, isolated binary evolution has been considered as a leading formation channel, in which the CHE channel was recently found to play an essential role in contributing to the whole population of BBH mergers \citep{2016MNRAS.458.2634M,2016A&A...588A..50M,2016MNRAS.460.3545D,2020MNRAS.499.5941D,2021MNRAS.505..663R,2021ApJ...910..152Z,2022A&A...657L...8B}. However, none of these events has been reported with a significantly high evidence for being formed through this channel. 

In this work, we search for candidates of merging BBHs originating from the CHE channel in GWTC-3, using Bayesian inference with astrophysically-predicted priors. Assuming GWTC-3 events originated from isolated binary evolution, we reanalyse a subsample of events using a suite of state-of-the-art models \citep{2020MNRAS.499.5941D,2021A&A...647A.153B} as released by \cite{2021ApJ...910..152Z}. After performing the Bayesian inference for the target events, we report strong evidence ($\ln\OR = 3.4$) for GW190517\_055101 being formed through the CHE channel. Under the assumption that the selected events in our subsample are all formed through one of the three channels considered in this work, we thus obtain the lower limits on the local merger rate density of these channels, $11.45 ~\mathrm{Gpc^{-3}~yr^{-1}}$ (CEE), $0.18 ~\mathrm{Gpc^{-3}~yr^{-1}}$ (CHE) and $0.63 ~\mathrm{Gpc^{-3}~yr^{-1}}$ (SMT) at $90\%$ credible level, respectively.

It is still a challenge to quantitatively predict the BBHs' properties due to uncertain physical processes involved in the single and/or binary evolution of massive stars \citep{2022ApJ...925...69B}. The upper limit of the stellar-mass BH predicted by (pulsational) pair-instability supernovae \citep[e.g.,][]{2017ApJ...836..244W,2019ApJ...887...53F,2019ApJ...882...36M} is still uncertain. Additionally, the models we adopt in this work assume efficient AM transport \citep{2002A&A...381..923S,2019ApJ...881L...1F} in the progenitor massive stars, which leads to forming first-born BHs with negligible spins \citep{2018A&A...616A..28Q,2019ApJ...881L...1F}. We note that BHs could obtain a slightly large spin ($\sim 0.1$) depending on the physics accounted for in the stellar models \citep{2020A&A...636A.104B}. \cite{2022arXiv220803129S} recently showed new supports for efficient internal AM transport for studying the spins of stripped B-type stars. However, this efficient mechanism could be challenged \citep{2022ApJ...924..129Q} with the detection of GW190403\_051519 \citep{2021arXiv210801045T}, under the assumption that this merger event was formed through the CEE channel. In the upcoming O4 Observing run of the LVK, more events like GW190517\_055101 are expected to be detected, and hence, this will allow to unravel the population properties of BBHs predicted by the CHE channel and further put stronger constraints on the physical processes in the evolution of massive stars in close binary systems.

\acknowledgments
We thank Ilya Mandel for helpful comments on the manuscript. We thank Zi-Qing Xia for useful discussion. YQ acknowledges the support from the Doctoral research start-up funding of Anhui Normal University. YZW is supported by NSFC (Grant No. 12203101). This work was supported by the National Natural Science Foundation of China (NSFC, Grant Nos. 11863003, 12003002, 12103003, 12173010, 12192220, 12192221) and by the Natural Science Foundation of Universities in Anhui Province (Grant No. KJ2021A0106). SSB is supported by the Swiss National Science Foundation (project number PP00P2\_176868). GM has received funding from the European Research Council (ERC) under the European Union's Horizon 2020 research and innovation programme (grant agreement No 833925, project STAREX). DMW is supported by NSFC (No. 12073080, 11933010, 11921003) and the Chinese Academy of Sciences via the Key Research Program of Frontier Sciences (No. QYZDJSSW-SYS024).

All figures are made with the free Python module Matplotlib \citep{Hunter2007}. We would also like to thank all of the essential workers who put their health at risk during the COVID-19 pandemic, without whom we would not have been able to complete this work.

\clearpage

\bibliography{ref}{}
\bibliographystyle{aasjournal}

\appendix
\section{supplementary figure}\label{sec:sfig}

\begin{figure*}[htbp!]
     \centering
     \includegraphics[width=\textwidth]{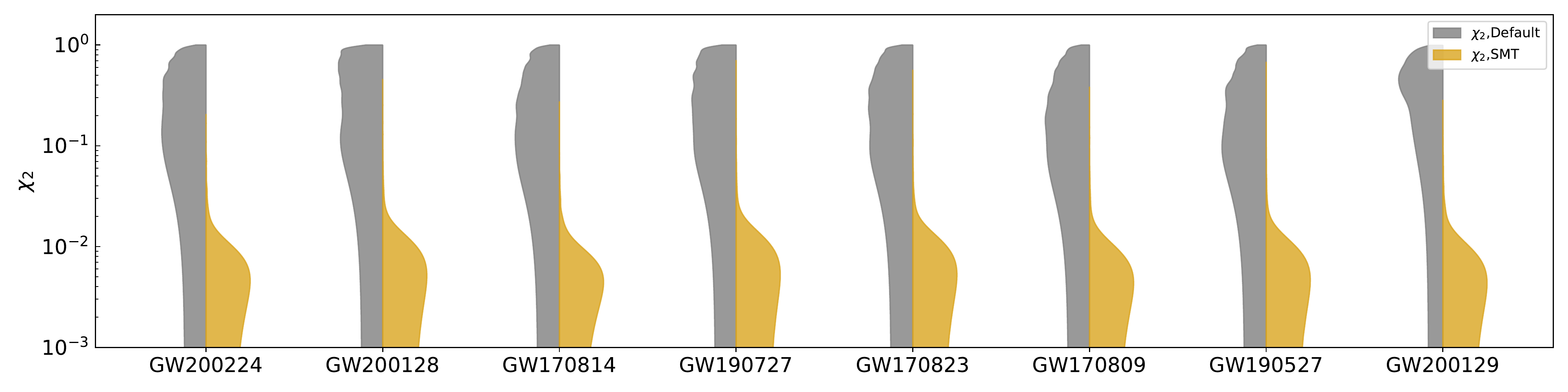}
     \includegraphics[width=\textwidth]{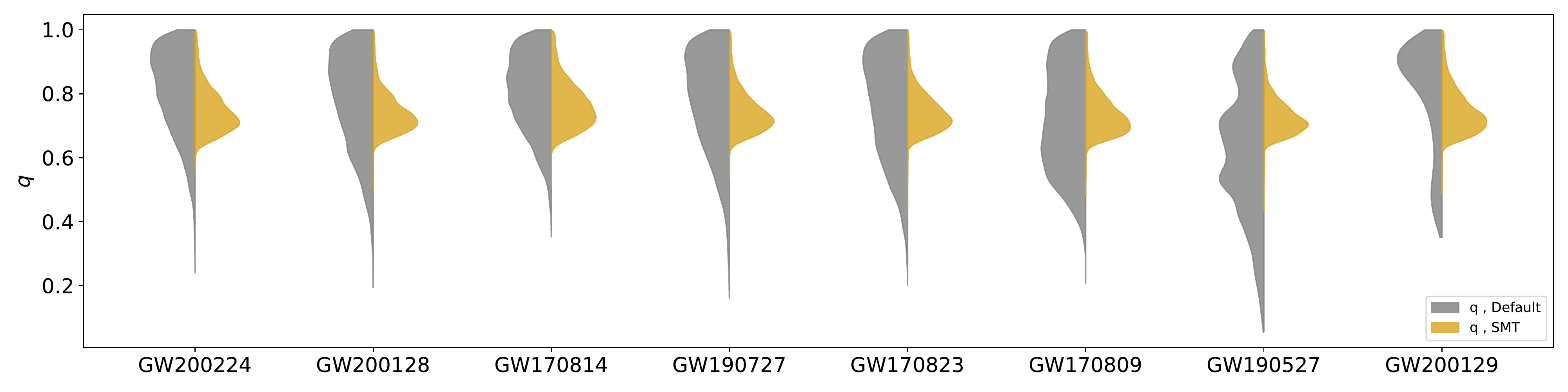}
     \caption{The same as Fig.~\ref{fig:violin}, but for the results inferred with the SMT prior. The selected events have odd ratios larger than $3$ for the SMT channel.}
     \label{fig:smt}
\end{figure*}

\end{document}